\documentclass[11pt]{article}

\usepackage{graphicx}

\begin{document}
\vbox {\vspace*{13mm}}
{\flushleft \LARGE {\bf Mid-IR-laser microscopy as a tool for defect\\
investigation in bulk semiconductors}}\\

\vbox {\vspace*{8mm}}
{\flushleft \Large {\bf O V Astaf$\,$iev, V P Kalinushkin and V A Yuryev}}\\
\\
General Physics Institute of RAS, 38,  Vavilov  Street,  Moscow,
117942, Russia\\
\\
\\
\hspace*{4mm} 
{\parbox{153mm}
{{\bf ABSTRACT: }
A non-destructive optical technique described in
this paper is an effective new tool  for  the investigation  of
defects  in  semiconductors.  The  basic  instrument  for   this
technique$\,$---$\,$a  mid-IR-laser  microscope$\,$---$\,$being sensitive  to
accumulations  of  free   carriers   enables  the  study of  both
accumulations of electrically-active defects  or  impurities  in
bulk semiconductors and doped domains in semiconductor
structures. The optical beam induced  scattering  mode  of  this
microscope is designed for the investigation of recombination-active
defects but unlike EBIC it requires neither Schottky barrier  or
$p$--$n$ junction nor special preparation of samples.}}\\
\\
\\
{\flushleft \bf 1. INTRODUCTION}\\

   At present, many techniques have been developed to study
defects on the micron and sub-micron scale in semiconductors.
Unfortunately most of them either break down the studied sample
or require rather complicated procedures of sample preparation
that in some cases may modify the properties of the defects
studied.  Moreover, practically all of the modern methods for
investigating such defects enable at best to observe defects but
cannot give information about their composition.  The
specificity of the techniques which may be applicable for
analysis of defect composition, such as e.g.  X-ray
microanalysis, microcathodoluminescence or photoluminescence,
does not allow one to investigate defects situated in crystal
bulk. Even the defects located in near-surface regions are hard
to be studied by these methods because of insufficient
sensitivity or, like in the case of X-ray microanalysis, because
the defects are often composed of intrinsic defects. So despite
the great success that now has been achieved in the
understanding of defect formation mechanisms, their composition
and properties, new methods for defect investigations, which
might appreciably accelerate the progress in these issues, are
presently required like they never were in the past.
Non-destructive methods for investigation of defects in crystal
bulk might e.g. give opportunities for direct studies of the
processes of inner gettering especially if they are combined
with techniques for studies of defects located in near-surface
layers. Considerable progress may be achieved in the
investigation of defects located in the vicinity of
semiconductor interfaces and surfaces coated with dielectric
layers and so on. We could give many examples of other
applications of such techniques (see e.g. Kalinushkin {\em et
al} 1995b). From our viewpoint, the most attractive methods for
such investigations are those which enables the study of the
crystal domains enriched with electrically-active defects and/or
defects interacting with non-equilibrium carriers (we call them
large-scale electrically-active defect accumulations, LSDAs, and
large-scale recombination-active defects, LSRDs).  The most
appropriate methods for investigating these defects are those
sensitive directly to micron-scale variations of the free
carrier concentration. This is because many of the procedures
for determining the point-defect structure of materials, e.g.
such as measurements of temperature dependencies of conductivity
or photocurrent measurements, may be suited for these
techniques.

   It seems, however, that until recently, the low-angle
mid-IR-laser light scattering technique, LALS (see Kalinushkin
1988, Voronkov {\em et al} 1990, Kalinushkin {\em et al} 1995b
and many references therein), was the only method that satisfied
all these requirements, and many studies of LSDAs and LSRDs in
different materials were carried out by this method~---~see e.g.
references cited by Kalinushkin {\em et al} (1995b) and
Kalinushkin (1988).  Nonetheless LALS in its standard form (or
LALS with angular resolution) has two main disadvantages:
firstly, it does not permit the study of each individual defect
and gives information averaged over a group of defects with
close parameters which are located within the probe beam, and
secondly, using only LALS one cannot estimate the concentrations
of LSDAs which are necessary to evaluate such parameters as free
carrier concentrations in LSDAs (and, hence, the energy
locations of point centers constituting them).  Therefore one
must use other methods to evaluate the concentrations of LSDAs
which decreases the reliability of the information obtained
(however, the data obtained e.g. by Kalinushkin {\em et al}
(1991) for InP and GaAs were confirmed by Yuryev and Kalinushkin
(1995) and Yuryev {\em et al} (1995) by direct observations).

   To overcome the mentioned shortcomings of LALS, a new method
for visualization of free carrier accumulations was recently
proposed (Astafiev {\em et al} 1994b, Kalinushkin {\em et al}
1995b, Astafiev {\em et al} 1995a, Astafiev {\em et al} 1995b
and Yuryev {\em et al} 1995), which is a kind of scanning laser
microscopy working in the mid-infrared wavelength range.  Being
an evolution of LALS (it is often referred to as scanning LALS
or SLALS), this method possesses all its merits and admits all
modifications which make LALS so convenient for investigation of
LSDAs.  Before long, the optical beam induced LALS (OLALS) mode
of the mid-IR-laser microscopy was developed (Astafiev {\em et
al} 1995c, Astafiev {\em et al} 1995d and Kalinushkin {\em et
al} 1995b) on the basis of LALS with surface photoexcitation
(Kalinushkin {\em et al} 1994a and Kalinushkin {\em et al}
1995a) for investigation of LSRDs in near-surface and
near-interface layers of semiconductors. The latter technique is
a non-destructive contactless optical analog of EBIC or, more
exactly, OBIC but in contrast to these methods OLALS does not
require Schottky barrier or $p$--$n$ junction and any special
surface preparation.

   The present paper is devoted to the description of the
developed techniques of mid-IR-laser microscopy and illustration
of their serviceability by the experimental results obtained for
bulk Si single crystals and their near-surface regions.\\

{\flushleft {\bf 2. SLALS MODE OF MID-IR-LASER MICROSCOPE}}\\
{\flushleft {\bf 2.1 Basic Instrument}}\\

   The SLALS technique is shown schematically in Fig.1. The
plane wave of the laser source illuminates a thin semiconductor
parallel-sided crystal with polished surfaces (usually standard
technological wafer before structure production). The wafer is
located in the focal plane of a lens L1. Let a defect be in the
crystal bulk in the front focal plane of L1.  It scatters the
probe wave producing an additional scattered wave, which
diverges in an angle of an order of $\lambda /a$ where $\lambda$
is the wave-length, {\it a} is the defect's characteristic size.
A resultant wave after the defect is the sum of the undisturbed
plane wave and that scattered by the defect. The lens L1
condenses the plane wave in the back focus to a spot with the
size of about $\lambda f_1/D1$ where $D1$ is the diameter of the
probe plane-wave beam, $f_1$ is the focal length of the lens L1.
A small mirror turned to the angle of 45$^{\circ}$ to the focal
plane or an absorbing screen is positioned in the back focal
plane to remove the probe wave radiation. The scattered wave,
being almost a plane wave with characteristic beam diameter of
$\lambda f_1/a$ after L1, passes to the second lens L2 almost
without losses if the screen size is smaller than $\lambda f_1/a$.
So the scattered wave without probe radiation reaches
the lens L2 and the image of the defect is formed in the back
focal plane of L2 in scattered rays.

\begin{figure}[t]
%\vspace*{35mm}
\begin{center}
%\hspace*{9.5mm}
 \includegraphics[scale=4]{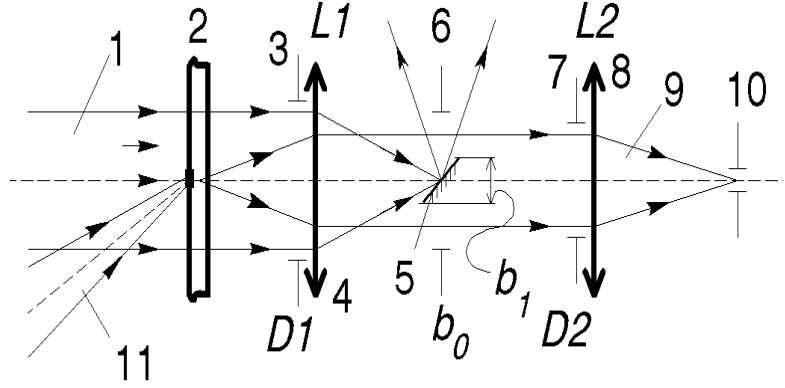}
 \end{center}
\caption{Optical diagram of the central dark ground
mid-IR-laser microscope:
(1) mid-IR probe wave; (2) sample; (3,6,7) diaphragms; (4,8) lenses;
(5) mirror or opaque screen; (9) scattered wave; (10) IR photodetector;
(11) exciting light beam (used in OLALS).}
\end{figure}

   The detailed analysis of the SLALS technique with application
to the the central dark field method of spatial-frequency signal
filtering, which is schematically presented in Fig.1, is made in
the paper by Astafiev {\em et al} (1995b).\\

{\flushleft \bf 2.2 LSDAs in CZ Si:B}\\

   The single crystals of standard industrial Si:B studied in
this work were grown by the Czochralski (CZ) process and had the
specific resistivity of 12~$\Omega \cdot$cm. The thickness of
the samples was 300 $\mu$m.  The large-scale electrically active
defects with the dimensions ranging from 3--5~$\mu$m to 40--50
$\mu$m were observed previously in analogous crystals in the
works by Buzynin {\em et al} (1990) and Astafiev et al (1994a).
Their concentrations were estimated as 10$^5$--10$^7$~cm$^{-3}$,
the values of the relative variations of the dielectric
constants $\delta \epsilon _m$ in them were evaluated as
10$^{-3}$--10$^{-4}$.

   The specimens sharply differing in the concentration of
defects as determined by selective etching were studied in the
present work: their concentrations were about
2$\times$10$^5$~cm$^{-3}$ in the sample 1 and about
2$\times$10$^4$~cm$^{-3}$ in the sample 2.

\begin{figure}[h]
%\vspace*{53mm}
\begin{center}
%\hspace*{9.5mm}
 \includegraphics[scale=1.2]{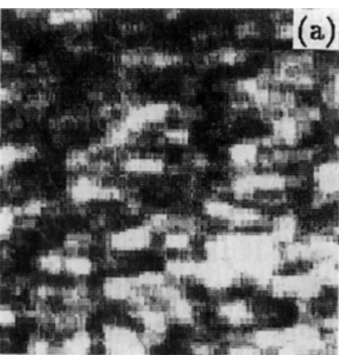}
\includegraphics[scale=1.2]{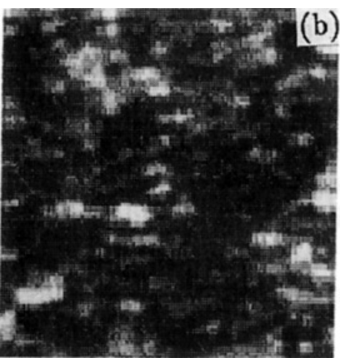}
\includegraphics[scale=1.2]{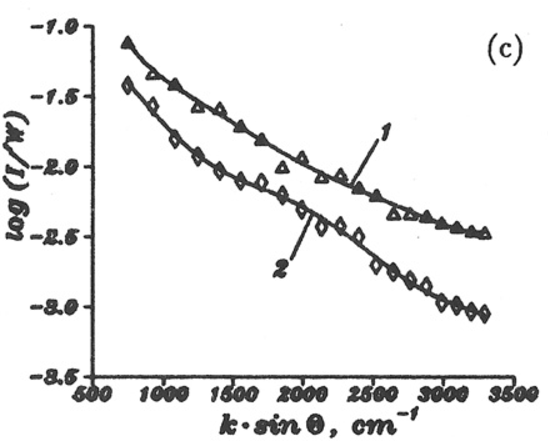}
 \end{center}
\caption{LSDAs in CZ Si:B (left to right): SLALS images
of the samples 1 $(a)$
and 2 $(b)$, 1$\times$1~mm$^2$; LALS diagrams $(c)$ for the samples
1 (1) and 2 (2); $\lambda$=10.6~$\mu$m.}
\end{figure}

   Fig.2$(a,b)$ presents the microphotographs of these samples
obtained with the use of SLALS (the areas of 1$\times$1mm$^2$
are depicted in the photographs). Like in the photographs of
indium phosphide and gallium arsenide (Yuryev and Kalinushkin
1995), the white spots in the pictures are the images of LSDAs.
The mean value of the light scattering intensity by the defects
in Fig.2$(a)$ is by around 3 times greater than that in
Fig.2$(b)$, that is completely in agreement with the results
obtained by LALS~---~the ratio of the light scattering
intensities by the samples 1 and 2 in the LALS diagrams given in
Fig.2$(c)$ is also nearly equal to 3.  The concentration of
defects determined in the sample 1 was about 10$^6$~cm$^{-3}$,
whereas that in the sample 2 was about
(4--5)$\times$10$^5$~cm$^{-3}$.

   Note that the values of the defect concentrations revealed in
the analogous samples by EBIC with the special surface
preparation (Buzynin {\em et al} 1989) appeared to be nearly
equal to the above values obtained by SLALS.

   The results obtained allow us to conclude the following: 1)
using the SLALS-microscope we visualized the defects which were
previously investigated by LALS$\,$---$\,$the so called "weak
impurity accumulations" (Buzynin {\em et al} 1990 and Astafiev
{\em et al} 1994a); 2) using EBIC with the special surface
preparation$\,$---$\,$the data obtained by this method were used
by Buzynin {\em et al} (1990) and Astafiev {\em et al}
(1994a)$\,$---$\,$we also revealed namely "weak impurity
accumulations", i.e. the estimations of LSDAs parameters and
thermal activation energies of the centers constituting LSDAs
made by Buzynin {\em et al} (1990) and Astafiev et al (1994a)
are valid.

   The radii of the accumulations calculated from the LALS
diagrams given in Fig.2$(c)$ are $a \sim \, 10-14 \, \mu$m.
Measuring the zero-angle light scattering intensity I$_0$/W from
LALS and determining the concentration of LSDAs from SLALS one
can easily estimate the values of $\delta
\epsilon _m$ in LSDAs.  They  are (6.5--10)$\times$10$^{-4}$  and
(11--14)$\times$10$^{-4}$ for the samples 1 and 2, respectively.

Note also that the correlation between the concentrations  of  defects
revealed by selective etching and SLALS is purely qualitative.  This  is
also characteristic for  most  of  the  comparative  experiments  on
etching  and  LALS.  However,  enquiring  into  the   reasons for the
discrepancies observed is beyond the scope of this paper.\\

{\flushleft \bf 3. OPTICAL BEAM INDUCED LALS}\\
{\flushleft \bf 3.1 Description of the Method}\\

   This mode of SLALS was recently proposed by us (Kalinushkin
{\em et al} 1995b and Astafiev {\em et al} 1995c, 1995d) as a
scanning modification of LALS with sample photoexcitation
(Kalinushkin 1988, Kalinushkin {\em et al} 1994 and Kalinushkin
{\em et al} 1995a). Its principle diagram is given in Fig.1.
The method, as well as LALS, can work in two modes: in the mode
of bulk photoexcitation and in the mode of surface
photoexcitation.  The regimes are different only by the choice
of pumping laser: the first regime uses a laser with quantum
energy less than the studied sample bandgap, whereas the second
one uses a laser with quantum energy greater than the bandgap,
and in general both regimes are quite analogous.

   The essence of the method, say for surface excitation, is as
follows. A highly focused beam (in contrast to LALS with
photoexcitation where a wide beam is used) generates
electron-hole pairs in the sample (in the chosen case, in its
near-surface region).  If the characteristic dimensions of the
non-uniformity of the generated electron-hole pair distribution
are as small as it is required in the paper by Astafiev {\em et
al} (1995b), the scattered mid-IR-laser light of the SLALS
microscope starts reaching the photodetector.  Its intensity is
proportional to the square of generated carrier concentration in
the spot. The characteristic sizes of the non-uniformity are
controlled by the sizes of the exciting laser spot, the carrier
diffusion length and the surface recombination velocity.  Even
if the diffusion length is large (e.g. in Si), the inhomogeneity
with small enough characteristic dimensions remains in the
carrier distribution because of the surface recombination.  This
inhomogeneity is detected by the SLALS microscope.  The carrier
concentration in such a "droplet" is controlled with the
electron-hole pair life-time in a given area of the sample.

   It is clear that the method is a very close analog of the
electron or optical beam induced current (EBIC or OBIC) methods,
but is different in that OLALS requires neither a Schottky
barrier nor a $p$--$n$ junction. It also does not require any
special preparation of surfaces.

   But the most important property of the developed method is
its ability to obtain information from interfaces and surfaces
covered with coatings and epilayers until the wafer is
metallized.

   It is also possible to create a kind of tomographic
microscopy on the basis of OLALS, and this problem does not seem
to be very difficult.

   Note that modulated 50 mW He-Ne laser radiation at the
wavelength of 0.63~$\mu$m was used in this work in the OLALS
experiments. The signal was detected with lock-in nanovoltmeter
at the modulation frequency.\\

\begin{figure}[t]
%\vspace*{47mm}
\begin{center}
%\hspace*{9.5mm}
 \includegraphics[scale=1.3]{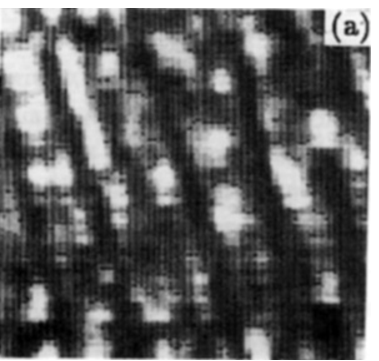}
\includegraphics[scale=1.3]{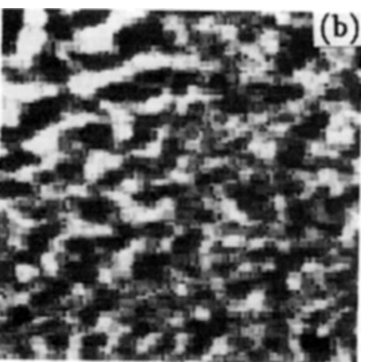}
\includegraphics[scale=1.3]{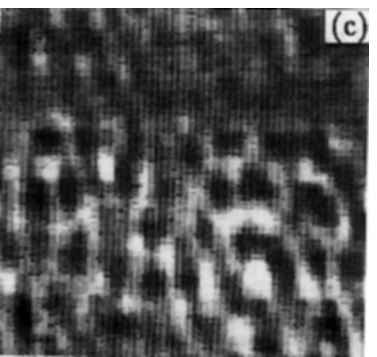}
 \end{center}
\caption{OLALS images, left to right: FZ Si:P,
chemico-dynamic $(a)$ and
mechanical $(b)$ polishing; CZ Si:B coated with 1200 {\AA} thick SiO$_2$
film $(c)$; 1$\times 1 \,$mm$^2$, $\lambda =~10.6\, \mu$m,
${\lambda}_{ex}=~633\,$nm.}
\end{figure}

\begin{figure}
%\vspace*{53mm}
\begin{center}
%\hspace*{9.5mm}
 \includegraphics[scale=4.5]{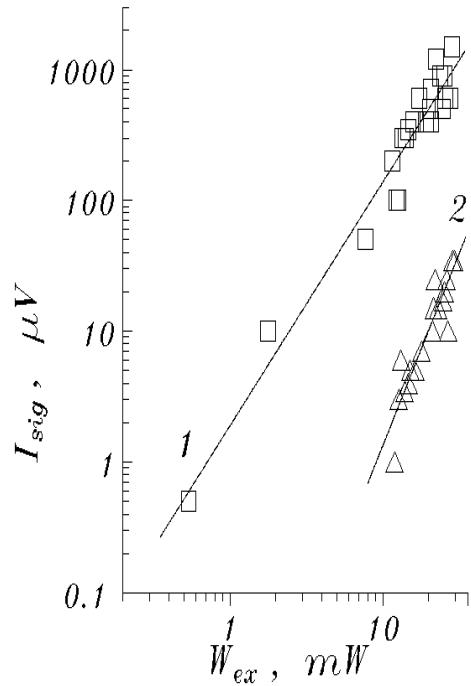}
 \end{center}
\caption{Dependences of MCT photodetector signal on the
absorbed power of
He-Ne laser radiation for chemico-dynamically (1) and mechanically (2)
polished sides of FZ Si:P wafer depicted in Fig.3; $\lambda =~10.6\, \mu$m,
${\lambda}_{ex}=~633\,$nm.}
\end{figure}

{\flushleft \bf 3.2 OLALS Images of Silicon Wafers}\\

   Fig.3 demonstrates OLALS images of FZ Si:P wafer surfaces
(1$\times$1mm$^2$). The pictures $(a)$ and $(b)$ present two
sides of the same wafer, one of which was polished in
chemico-dynamic way ("finished side") and the other was polished
mechanically up to optical precision grade.  The darker the
areas in the photographs, the shorter the carrier life is.  So
the dark stripes in the pictures of the finished side
correspond, from our viewpoint, to tracks of underpolished
and/or underetched scratches.  A very badly damaged layer is
registered on the mechanically polished side (the signal from
this side was by around 1000 times lower than that from the
finished side). The picture $(c)$ gives the micrograph of CZ
Si:B wafer surface coated with 1200 {\AA} thick layer of SiO$_2$
(this wafer was taken directly from the technological line of
CCD production). The dark spots are likely the images of
defective regions.

   Fig.4 demonstrates the dependence of IR-photodetector signal
on the power of the exciting He-Ne laser for the FZ Si:P wafer
depicted in Fig.3.  Two cases are shown: for the finished side
(marked as 1) and for the mechanically polished side (marked as
2). It is seen that in the case (1) the signal is proportional
to the square of the photoexciting laser intensity whereas in
the case (2) the signal is proportional to the third power of
the He-Ne laser intensity.  These lines confirm that the SLALS
microscope works with scattered rays and the scattering by the
domain with generated non-equilibrium carriers allows us to make
imaging in the OLALS mode. The cubic dependence (2) has not been
explained yet. Note that the same depence was obtained for the
mechanically polished Ge sample in the work by Kalinushkin {\em
et al} (1995a).\\

{\flushleft \bf 4. CONCLUSION}\\

   It was shown in this paper that the mid-IR-laser microscopy
may become a useful tool for defect investigations in
semiconductors. This method can be easily complemented with such
well developed techniques for defect composition analysis as
measurements of temperature dependencies of LALS (see e.g.
Kalinushkin {\em et al} 1995 and Kalinushkin {\em et al} 1991).
The measurements of LALS photoexcitation spectra might be also
very useful for this purpose. Such well known technique as
photoluminescence mapping might be used as a complementary
method to the OLALS mode, with the same exciting laser being
used for both techniques and the measurements being made
simultaneously. The tomographic SLALS microscope is also under
development now using the principles of laser heterodyning
(Sawatari 1973 and Protopopov and Ustinov 1985). The methods of
phase contrast and interference microscopies (Fran\c{c}on 1954)
also might appear to be useful.\\

{\flushleft \bf REFERENCES}\\
{\flushleft Astafiev O  V, Buzynin  A  N,  Buvaltsev  A  I  {\em et al}
1994a  Semicond. \underline{28} (3) 407}\\
Astafiev O V, Kalinushkin V P and Yuryev V A 1994b Proc. SPIE
\underline{2332} 138\\
Astafiev O V, Kalinushkin V P and Yuryev V A 1995a Mater. Sci. Eng. (B)
in press\\
Astafiev O V, Kalinushkin V P and  Yuryev  V  A  1995b  Microelectronics
in press\\
Astafiev O V, Kalinushkin V P and Yuryev V A 1995c Microelectronics
in press\\
Astafiev O V, Kalinushkin V P and Yuryev V A 1995d J. Tech. Phys. Lett.
in press\\
Buzynin A N, Butylkina N A, Kalinushkin V  P  {\em et al}  1989  USSR  Patent
No.1531766\\
Buzynin A N, Zabolotskiy S E, Kalinushkin V P {\em et  al} 1990
Sov. Phys.--$\,$Semicond. \underline{24} (2) 264\\
Fran\c{c}on M\   1954\  Le\  Microscope\  \'{a} \ Contraste \ de\  Phase\  et\  Microscope\
Interferentiel \ (Paris:\\
\hbox {\hspace*{5mm}}{Centre National de la Recherche Scientifique)}\\
Kalinushkin  V  P  1988  Proc. Inst. Gen. Phys. Acad. Sci. USSR  Vol.4 Laser
Methods of Defect\\
\hbox {\hspace*{5mm}}{Investigations in Semiconductors and Dielectrics (New
York: Nova) pp.1--75}\\
Kalinushkin V P, Murin D I, Astafiev O V {\em et  al} 1995a Phys.
Chem. Mech. Surf. (4) in press\\
Kalinushkin V P, Murin D I, Yuryev V A {\em et  al} 1994 Proc. SPIE
\underline{2332} 146\\
Kalinushkin V P, Yuryev V A and Astafiev O V  1995b  Mater. Sci. Technol.
in press\\
Kalinushkin V P, Yuryev V A, Murin D I {\em et  al} 1991 Semicond. Sci.
Technol. \underline{7} A255\\
Protopopov V V and Ustinov N D 1985 Laser Heterodyning (Moscow: Nauka)\\
Sawatari T 1973 Appl.Opt. \underline{12} 2768\\
Voronkov V V, Zabolotskiy S E, Kalinushkin V P {\em et  al} 1990 J. Cryst.
Growth \underline{103} 126\\
Yuryev V A, Astafiev O V and Kalinushkin V P 1995 Semicond. in press\\
Yuryev V A and Kalinushkin V P 1995 Mater.Sci.Eng. (B) in press
\end{document}